\documentclass[twocolumn]{aastex6}

\slugcomment{Accepted to the Astrophysical Journal }

\newcommand\msun{\ensuremath{\rm M_{\odot}}}

\newcommand\kms{\ensuremath{\rm km\, s^{-1}\;}}

\newcommand\be {\begin{equation}}
\newcommand\ee{\end{equation}}
\newcommand\fesixty{\ensuremath{\rm ^{60}Fe\ }}
\newcommand\fesixtyns{\ensuremath{\rm ^{60}Fe}}

\newcommand\altwentysix{\ensuremath{\rm ^{26}Al\ }}
\newcommand\altwentysixns{\ensuremath{\rm ^{26}Al}}
\newcommand\altwentyseven{\ensuremath{\rm ^{27}Al\ }}
\newcommand {\gtsim} {\ {\raise-.5ex\hbox{$\buildrel>\over\sim$}}\ }
\newcommand {\ltsim} {\ {\raise-.5ex\hbox{$\buildrel<\over\sim$}}\ }

\begin{document}


\title{Triggered Star Formation Inside the Shell of a Wolf-Rayet Bubble as the
  origin of the Solar System }


\author{Vikram V. Dwarkadas} 
\affil{Astronomy and Astrophysics, University of Chicago, 5640 S Ellis Ave, ERC 569, Chicago, IL 60637}
\email{vikram@oddjob.uchicago.edu}
\author{Nicolas Dauphas}
\affil{Origins Laboratory, Department of the Geophysical Sciences and Enrico Fermi Institute, The University of Chicago, 5734 South Ellis Avenue, Chicago IL 60637}
\email{dauphas@uchicago.edu}
\author{Bradley Meyer}
\affil{Department of Physics and Astronomy, Clemson University, Clemson, SC, 29634-0978, USA. }
\email{mbradle@g.clemson.edu}
\author{Peter Boyajian}
\affil{Astronomy and Astrophysics, University of Chicago, 5640 S Ellis Ave, ERC 569, Chicago, IL 60637}
\email{peterhaigboyajian@gmail.com}
\and
\author{Michael Bojazi}
\affil{Department of Physics and Astronomy, Clemson University, Clemson, SC, 29634-0978, USA. }
\email{mbojazi@g.clemson.edu}



\begin{abstract}
  A critical constraint on solar system formation is the high
  \altwentysix/$^{27}$Al abundance ratio of 5 $\times 10^{-5}$ at the
  time of formation, which was about 17 times higher than the average
  Galactic ratio, while the $^{60}$Fe/$^{56}$Fe value was about $2
  \times 10^{-8}$, lower than the Galactic value. This challenges the
  assumption that a nearby supernova was responsible for the injection
  of these short-lived radionuclides into the early solar system. We
  show that this conundrum can be resolved if the Solar System was
  formed by triggered star formation at the edge of a Wolf-Rayet (W-R)
  bubble.  Aluminium-26 is produced during the evolution of the
  massive star, released in the wind during the W-R phase, and
  condenses into dust grains that are seen around W-R stars. The dust
  grains survive passage through the reverse shock and the low density
  shocked wind, reach the dense shell swept-up by the bubble, detach
  from the decelerated wind and are injected into the shell. Some
  portions of this shell subsequently collapses to form the dense
  cores that give rise to solar-type systems. The subsequent
  aspherical supernova does not inject appreciable amounts of \fesixty
  into the proto-solar-system, thus accounting for the observed low
  abundance of \fesixtyns. We discuss the details of various processes
  within the model and conclude that it is a viable model that can
  explain the initial abundances of \altwentysix and \fesixtyns. We
  estimate that 1-16\% of all Sun-like stars could have formed in such
  a setting of triggered star formation in the shell of a WR bubble.
\end{abstract}




\section{Introduction} 
\label{sec:intro}

The discovery and subsequent characterization of extrasolar planetary
systems has shed new light on the origin and peculiarities of the
solar system. Astronomical observations offer clues about the grand
architecture of planetary systems. They cannot provide, however,
access to the intricate details that the study of meteorites
offers. One important constraint on models of solar system formation
comes from measurements of the abundances of now extinct short-lived
radionuclides, whose past presence is inferred from isotopic
variations in their decay products. Isotopic abundances in meteorites
provide insight into the makeup of the cloud material from which the
solar system formed. They can be used as tracers of the stellar
processes that were involved in the formation of the solar system, and
of galactic chemical evolution up until the time of solar system
formation.

More than 60 years ago, \citet{urey55} speculated about the possible
role of \altwentysix as a heat source in planetary bodies. It was not
until 1976, however, that \citet{lpw76} demonstrated the presence of
this radioactive nuclide in meteorites at a high level. The high
abundance of $^{26}$Al (${\rm ^{26}Al/^{27}Al\sim 5\times 10^{-5}}$)
at solar system birth \citep{lpw76,mdz95,jacobsenetal08} can be
compared to expectations derived from modeling the chemical evolution
of the Galaxy \citep{mc00,wasserburgetal06,hussetal09}, or
$\gamma$-ray observations \citep{diehletal06}, which give a maximum
${\rm ^{26}Al/^{27}Al}$ ratio of $\sim 3\times 10^{-6}$
\citep{td12}. Aluminium-26 in meteorites is in too high abundance to
be accounted for by long-term chemical evolution of the Galaxy or
early solar system particle irradiation \citep{mgd02,dt07}. Instead,
\altwentysix must have been directly injected by a nearby source (see
\S \ref{sec:altsix}). Such sources can include supernovae
\citep{ct77,mc00}, stellar winds from massive stars \citep{apm97,
  diehletal06, agm06, gaidosetal09, gm12, young14, td15, young16}, or
winds from an AGB-star \citep{wasserburgetal06}. The latter is
unlikely, because of the remote probability of finding an evolved star
at the time and place of solar system formation \citep{km94}.  Recent
calculations by \citet{wkl17} have shown that it is unlikely that an
AGB star could simultaneously account for the abundances of
\altwentysixns, \fesixtyns, $^{182}{\rm Hf}$ and $^{107}{\rm
  Pd}$. Aluminium-26 is mainly produced by hot bottom burning in stars
$\ga 5 \msun$, which produce too little $^{182}{\rm Hf}$ and
$^{107}{\rm Pd}$. The latter are mainly products of neutron capture
processes in stars $ \la 5 \,\msun$.  A small window of AGB star
masses between 4-5.5 \,\msun~could be made to work, but this requires
that hot bottom burning in these stars was stronger than was assumed
in their calculations, and/or an additional neutron source was
present.

One way to test whether supernovae or stellar winds from massive stars
are the source is to examine \fesixty ($t_{1/2}$=2.6 Myr)
\citep{wgb98}. This radioactive nuclide is produced mostly by neutron
capture in the inner part of massive stars, whereas \altwentysix is
produced in the external regions \citep{lc06}. If a supernova injected
\altwentysix, one would expect copious amounts of \fesixty to also be
present.

The formation of refractory Ca, Al-rich inclusions (CAIs) in
meteorites marks time zero in early solar system chronology
\citep{dc11}. To better constrain the ${\rm ^{60}Fe/^{56}Fe}$ ratio at
CAI formation, many studies in the early 2000's analyzed chondrites
using Secondary Ion Mass Spectrometry (SIMS) techniques, and reported
high ${\rm ^{60}Fe/^{56}Fe}$ ratios of $\sim {\rm ^{60}Fe/^{56}Fe = 6
  \times 10^{-7}}$. For reference, the expected ${\rm
  ^{60}Fe/^{56}Fe}$ ratio in the interstellar medium (ISM) 4.5 Ga from
$\gamma$-ray observations \citep{wangetal07} is $\sim 3\times 10
^{-7}$ without accounting for isolation from fresh nucleosynthetic
input before solar system formation \citep{td12}. The SIMS results
were later shown to suffer from statistical artifacts, leading
\citet{telusetal12} to revise downward the initial ${\rm
  ^{60}Fe/^{56}Fe}$ ratios that had been previously reported.

The high initial ${\rm ^{60}Fe/^{56}Fe}$ ratios inferred from {\it in
  situ} measurements of chondritic components contradicted lower
estimates obtained from measurements of bulk rocks and components of
achondrites. This led to speculations that $^{60}$Fe was
heterogeneously distributed in the protoplanetary disk, with
chondrites and achondrites characterized by high and low ${\rm
  ^{60}Fe/^{56}Fe}$ ratios, respectively
\citep{smy06,quitteetal10,quitteetal11}.

The issue of the abundance and distribution of $^{60}$Fe in meteorites
was addressed by Tang and Dauphas \citep{td12,td15}. Using Multi
Collector Inductively Coupled Plasma Mass Spectrometry (MC-ICPMS),
these authors measured various components from chondrites Semarkona
(LL3.0), NWA 5717 (ungrouped 3.05) and Gujba (CBa), as well as bulk
rocks and mineral separates from HED and angrite achondrites. They
showed that in these objects, the initial ${\rm ^{60}Fe/^{56}Fe}$
ratio was low, corresponding to an initial value at the formation of
CAIs of 11.57 $\pm$ 2.6 $\times 10^{-9}$. They also measured the ${\rm
  ^{58}Fe/^{56}Fe}$ ratio and found that it was constant between
chondrites and achondrites, thus ruling out a heterogeneous
distribution of \fesixty as collateral isotopic anomalies on $^{58}$Fe
would be expected. Some SIMS studies have continued to report higher
ratios but the data do not define clear isochrones
\citep{mishraetal10, mc12,mm12, mc14,mg14, telusetal16}. Furthermore,
recent measurements using the technique of Resonant Ionization Mass
Spectrometry (RIMS; \citep{trappitschetal17, boehnkeetal17} ) have
called into question the existence of the $^{60}{\rm Ni}$ excesses
measured by SIMS.  {\em The weight of evidence at the present time
  thus favors a low uniform initial ${\rm ^{60}Fe/^{56}Fe}$ ratio at
  solar system formation} \citep{td12,td15}.

The low initial ${\rm ^{60}Fe/^{56}Fe}$ ratio may be consistent with
derivation from background abundances in the Galaxy with no compelling
need to invoke late injection from a nearby star \citep{td12}.
Iron-60 is a secondary radioactive isotope and ${\rm ^{56}Fe}$ is a
primary isotope with respect to nucleosynthesis. To first order,
models of galactic chemical evolution find that the ratio of the
abundance of a primary nuclide (proportional to time $t$) to a
secondary nuclide (proportional to $t \times \tau$, since it depends
on the primary isotope, as well as on its half-life $\tau$), should be
roughly constant in time over Galactic history \citep{hussetal09}.
This indicates that the ISM ratio at the time of the Sun's birth was
probably near $\sim 3\times 10^{-7}$, the current value inferred from
$\gamma$-ray observations.  This value, however, is an average over
all phases of the ISM.  The ${\rm ^{60}Fe}$ ejected from supernovae
predominantly goes into hot material that then takes some tens of
millions of years to cool down to a phase that can undergo new star
formation.  We expect that this delay reduces the ${\rm ^{60}Fe}/{\rm
  ^{56}Fe}$ ratio to the low value inferred from chondrites.

The fundamental challenge in reconciling the early solar system
abundances of ${\rm ^{26}Al}$ and ${\rm ^{60}Fe}$ is thus to
understand how to incorporate freshly made $^{26}$Al without adding
too much $^{60}$Fe. Adjusting the timescale between nucleosynthesis
and solar system formation does not help because $^{26}$Al has a
shorter half-life than $^{60}$Fe, so any delay would cause the ${\rm
  ^{26}Al/^{60}Fe}$ ratio to decrease, making the problem worse. Some
possible scenarios that have been suggested to explain the high ${\rm
  ^{26}Al/^{60}Fe}$ ratio of the early solar system include (1)
supernova explosion with fallback of the inner layers, so that only
$^{26}$Al is efficiently ejected while $^{60}$Fe is trapped in the
stellar remnant \citep{mc00}, but this is unlikely because one would
need to have fallback that extends to external regions in the star (a
cutoff in the C$/$O-burning layer) to prevent $^{60}$Fe from escaping
\citep{takigawaetal08}, (2) interaction of a supernova with an already
formed cloud core, so that only the outer layers are efficiently
injected while the inner layers are deflected
\citep{gritschnederetal12}, and (3) ejection of $^{26}$Al in the winds
of one or several massive stars \citep{apm97, diehletal06, agm06,
  gaidosetal09, tdd10, gm12, young14, td15, young16}. The last
scenario is the most appealing because it could be a natural outcome
of the presence of one or several Wolf-Rayet (W-R) stars, which shed
their masses through winds rich in $^{26}$Al, whereas $^{60}$Fe is
ejected at a later time following supernova explosion. The W-R winds
expanding out into the surrounding medium would have carved wind-blown
bubbles in molecular cloud material, and would have enriched the
bubbles in $^{26}$Al, which would have later been incorporated in the
molecular cloud core that formed the solar system.

In this contribution, we take this idea a step further and suggest
that our solar system was born inside the shell of a Wolf-Rayet wind
bubble. Using a combination of semi-analytic calculations,
astronomical observations, and numerical modeling, we show that a
single massive star would produce enough \altwentysix to enrich the
entire solar system, that this \altwentysix would be incorporated into
the dense shell surrounding the wind-blown bubble, and that molecular
cores within the dense shell would later collapse to form the solar
protoplanetary nebula. In \S 2 we discuss the sources and yields of
\altwentysixns. \S 3 discusses the formation and evolution of W-R
bubbles, and the production of \altwentysix in these
bubbles. Triggered star formation at the periphery of these bubbles is
the topic of \S 4. The transport of \altwentysix is the subject of \S
5, including condensation of \altwentysix onto dust grains and
injection of \altwentysix into the dense shell. The subsequent SN
explosion, and whether the shell would be contaminated further by
explosive \fesixtyns, is discussed in \S 6. The timing of formation of
the proto-solar disk is discussed in \S 7. Finally, further discussion
of our model and conclusions are dealt with in \S 8.

\section{Sources of \altwentysix } 
\label{sec:altsix}
Aluminium-26 is a radioactive nuclide, with a half-life of about 0.7
Myr. The radioactive decay of \altwentysix leads to the emission of
1.8 MeV gamma-rays. More than 20 years ago, \citet{pd95}, discussing
Comptel data, suggested that ``massive stars embedded in the spiral
arms dominate the 1.8 MeV sky image''. In 1999, \citet{knodlseder99}
showed from an analysis of Comptel data that the 1.8 MeV gamma-ray
line was closely correlated with the 53 GHz free-free emission in the
Galaxy.  The free-free emission arises from the ionized interstellar
medium. They argued that this could be understood if massive stars are
the source of \altwentysixns, as had already been suggested by
\citet{pd96}. \citet{knodlseder99} showed that the correlation was
also strong with other tracers of the young stellar population.

The distribution of \altwentysix~in the Galaxy closely traces the
distribution of very massive stars, making W-R stars and core-collapse
SNe the primary candidates for \altwentysix~production. The former are
stars with initial mass \gtsim 25 \msun, which have lost their H and
possibly He envelopes. Many authors have suggested that stellar winds
from massive stars, could be the source of \altwentysix in the early
solar system \citep{apm97, diehletal06, agm06, gm12, young14, td15,
  young16}.  Aluminium-26 has been seen towards star forming regions
such as Vela \citep{oberlacketal94} and Cygnus
\citep{martinetal10}. \citet{vossetal10} used a census of the most
massive stars in Orion to compute the stellar content in the region,
followed by the ejection of \altwentysix from these stars. They found
good agreement between their model and the \altwentysix signal in the
Orion region. \citet{diehletal10} detected a significant \altwentysix
signal, $> 5 {\rm \sigma}$ above the background, from the
Scorpius-Centaurus region. In a study of the Carina region using
INTEGRAL data, \citet{vossetal12} found that the \altwentysix~signal
could not be accounted for by supernovae alone, and the fraction of
\altwentysix~ejected in W-R stars is high.

\subsection{Aluminium-26 in Massive Stars}
\label{sec:alyields}
Aluminium-26 is mainly produced in stars in the main sequence
H-burning phase, by the $^{25}{\rm Mg}(p,\gamma)\altwentysix$
proton-capture reaction \citep{lc06}. The conversion starts at the
beginning of the main sequence, and comes to completion within the
H-burning lifetime. The \altwentysix production reaches a maximum
shortly after the onset of H burning, after which it $\beta$ decays
into $^{26}{\rm Mg}$, on a timescale of order 10$^6$ yr. Since
production still continues, the \altwentysix abundance decrease is
slower than it would be for pure $\beta$ decay, but it does not reach
a stable state.  After the exhaustion of core H burning, the
\altwentysix is found in the He core and in the H left behind by the
receding core.  This would be mainly lost during the explosion, but
because massive stars have dredge ups and lose significant amounts of
mass, the \altwentysix can be expelled in the stellar wind. In stars
that become W-R stars, there are no significant dredge-up episodes,
\altwentysix is preserved in the He core, and ejected mainly through
mass loss which can reach deep into the interior \citep{lc06}. Thus,
although it is produced in the early stages, it is only in the
post-main-sequence phases, especially in W-R stars, that most of the
\altwentysix is expelled though wind loss, making W-R stars one of the
primary producers of \altwentysix into the interstellar
medium. Aluminium-26 is also produced in explosive nucleosynthesis,
but as we will show later this is inconsequential in our model.

\begin{figure}[htbp]
\includegraphics[width=\columnwidth, angle=90]{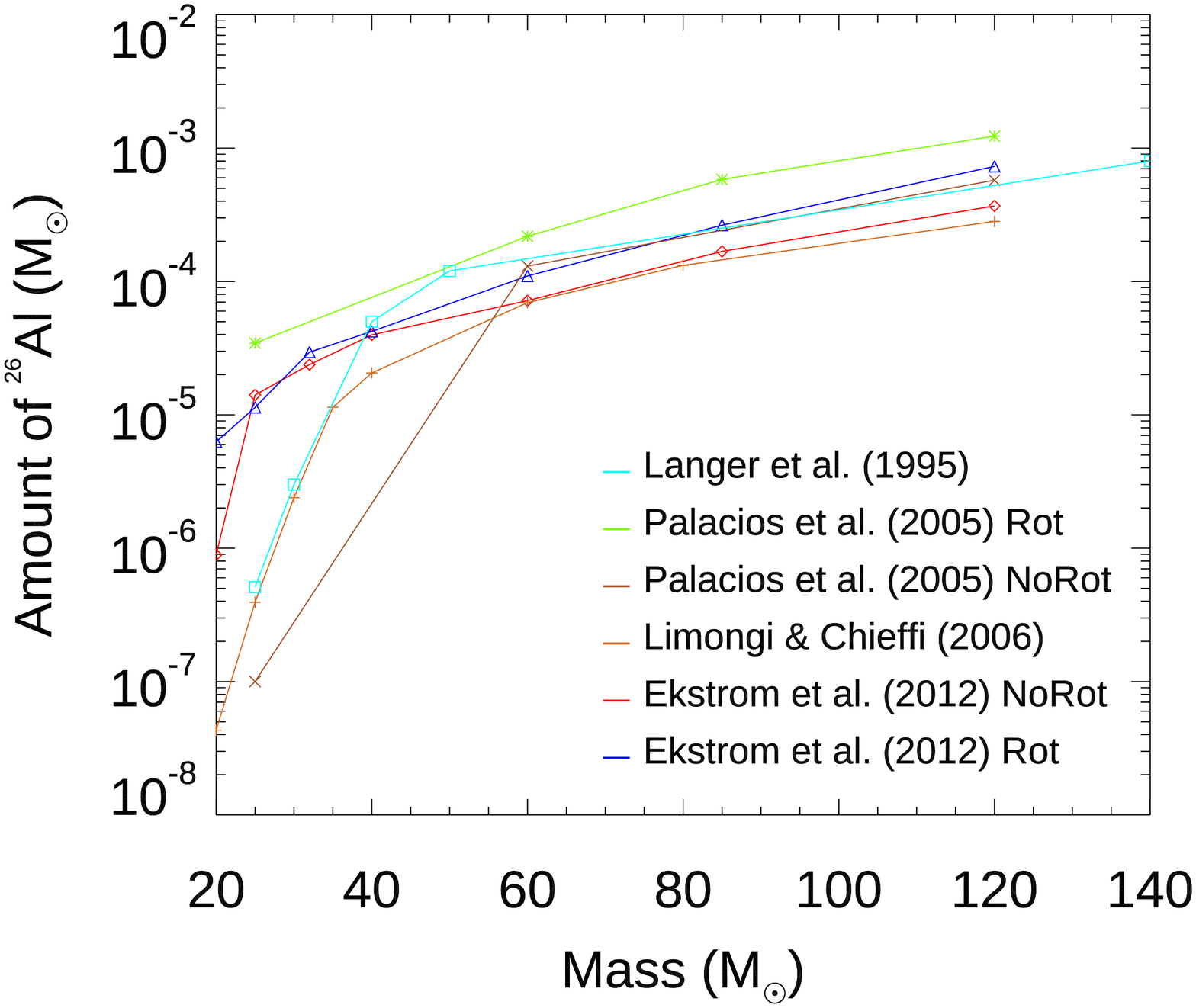}
\caption{Amount of \altwentysix lost from massive stars with mass $>
  20$ M$_{\odot}$.  ﻿References cited: \citet{ekstrometal12}: no
  rotation, Z=.014; \citet{ekstrometal12}: V/Vc =.4, Z=.014;
  \citet{lc06} only wind: no rotation; \citet{palaciosetal05}: Vo =0
  \kms, Z=0.02 ; \citet{palaciosetal05}: Vo =300 \kms, Z=0.02 ;
  \citet{lbf95} Wind, Z=.02. [Vc=critical rotation velocity (expressed in km s$^{-1}$, or as a fraction of critical velocity); Vo =
    surface rotation velocity]. }
\label{fig:yields}
\end{figure}

In order to quantify the production of \altwentysix in massive stars,
we have compiled computations of the \altwentysix production from
several groups \citep{lbf95, lc06, palaciosetal05, ekstrometal12,
  georgyetal12, georgyetal13}. Most of the models provide the total
\altwentysix yield at the end of the evolution of the star. The later
calculations \citep{ekstrometal12, georgyetal12, georgyetal13}, which
take into account updated solar metallicity (0.014), stellar rotation
and improved mass-loss rates, provide the \altwentysix yield
throughout the stellar evolution history, thus allowing us to evaluate
not only the total yield but also when \altwentysix was lost in the
wind, and thus take the decay of \altwentysix into account. In Figure
\ref{fig:yields} we have plotted the \altwentysix yields from stars
with initial mass $\ge$ 20 \msun.  In general, a single massive
Wolf-Rayet (W-R) star provides at least 10$^{-5}$~\msun ~of
\altwentysixns.  The more massive the star, the higher the
\altwentysix yield.  The spread in the \altwentysix amount produced is
due to differences in the nuclear reaction rates, which by itself can
produce a difference of up to a factor of 3 in the yield
\citep{iccl11}, as well as differences intrinsic to the stellar model
physics such as mixing, mass-loss rates, rotation velocities, and
magnetic fields. The \fesixty yield from the wind itself is
negligible, as the \fesixty is primarily produced in these massive
stars in the He convective shell, and is not ejected by the stellar
wind \citep{lc06}.

\section{Wolf-Rayet Bubbles}
W-R stars form the post-main- sequence phase of massive O-type main
sequence stars. The physical properties and plausible evolutionary
scenarios of W-R stars are detailed in \citet{crowther07}. Although
their evolutionary sequence is by no means well understood, it is
generally accepted that they form the final phase of massive stars $>
25 \msun$ before they core-collapse to form SNe.  These stars have
radiatively driven winds with terminal velocities of 1000-2000 km
s$^{-1}$ \citep{crowther01}, and mass-loss rates of order 10$^{-7}$ to
10$^{-5} \,\msun$ yr$^{-1}$ in the W-R phase.  The high surface
temperature of these hot stars ($>$ 30,000 K) results in a large
number of ionizing photons - the UV ionizing flux is of order
10$^{49}$ photons s$^{-1}$ \citep{crowther07}.

\begin{figure}[htbp]
\includegraphics[width=\columnwidth, angle=90]{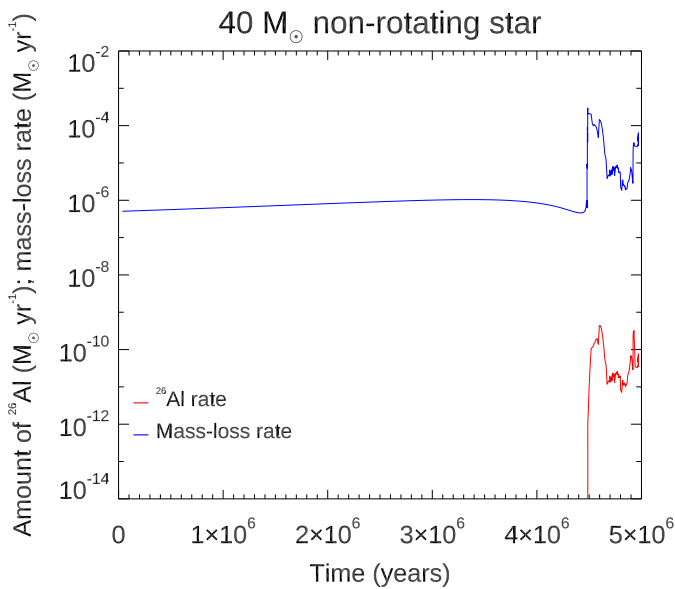}
\caption{The evolution of the wind mass-loss rate (blue) in a
  non-rotating 40 $\msun$ star. The parameters are taken from the
  stellar models of \citet{ekstrometal12}. The mass-loss rate is
  approximately steady or slowly increasing throughout the
  main-sequence phase, but increases dramatically in the
  post-main-sequence red supergiant and W-R phases. The figure shows
  (in red) the amount of \altwentysix lost per year via the wind.  }
\label{fig:massloss}
\end{figure}

In Figure \ref{fig:massloss}, we show the evolution of the wind
mass-loss rate in a 40 $\msun$ non-rotating star
\citep{ekstrometal12}. The mass-loss rate is lowest in the
main-sequence phase (up to $\sim$ 4.5 $\times 10^5$ years), increases
in the subsequent He-burning red supergiant phase, and then drops
somewhat in the W-R phase as the star loses its H envelope. The figure
also shows the \altwentysix loss rate (the \altwentysix emitted per
year) in the wind. Note that this also increases in the
post-main-sequence phases, and essentially follows the mass-loss.

The combined action of the supersonic winds and ionizing radiation
results in the formation of photo-ionized wind-blown bubbles around
the stars, consisting of a low-density interior surrounded by a
high-density shell.  The nature of wind-driven bubbles was first
elucidated by \citet{weaveretal77}. Going outwards in radius, one can
identify 5 distinct regions: (1) a freely expanding wind, (2) a
shocked wind region, (3) a photoionized region, (4) a thin dense
shell, and (5) the external medium. A reverse or wind termination
shock separates the freely expanding wind from the shocked wind. The
external boundary is generally a radiative shock, which compresses the
swept-up material to form a thin, dense shell, enclosed between the
radiative shock and a contact discontinuity on the inside.

\begin{figure}[htbp]
\includegraphics[width=\columnwidth]{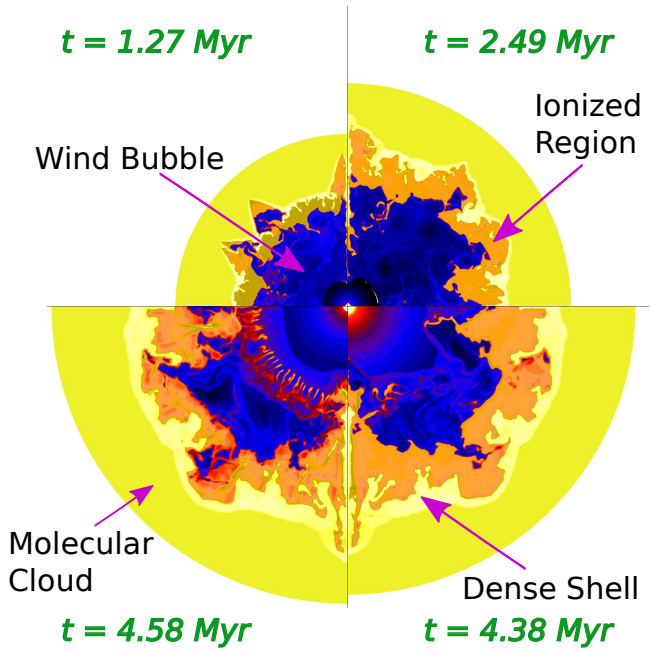}
\caption{Wind-Blown bubble around a Wolf-Rayet Star. The figure shows
  the density at 4 epochs in the evolution of a wind-blown bubble
  around a 40 $\msun$ star, starting clockwise from top left, at 1.27,
  2.49, 4.38 and 4.58 Myr. (The parameters for this simulation are
  taken from \citet{vanmarleetal05}, with mass-loss rates somewhat
  different from those shown in Figure \ref{fig:massloss}). The blue
  region is the wind-driven bubble, which is separated from the dense
  shell (light yellow) by the golden ionized region. The shell is
  unstable to several instabilities, related to both the hydrodynamics
  and the ionization front, which cause fragmentation and the
  formation of dense filaments and clumps. The wind-driven bubble in
  this case reaches the dense swept-up shell only during the
  Wolf-Rayet phase. }
\label{fig:wrbub}
\end{figure}

Models of wind-blown bubbles incorporating both the photo-ionizing
effects of the hot stars and the gas dynamics have been computed by
\citet{ta11} and \citet{dr13}. In Figure \ref{fig:wrbub} we show the
evolution of a W-R bubble around a 40 $\msun$ star.  Most of the
volume is occupied by the shocked wind region (blue in the figure)
which has a low density and high temperature. The ionizing photons
create the photo-ionized region which extends beyond the wind-blown
region in this particular case due to the high number of ionizing
photons. The thin shell is susceptible to various hydrodynamical
instabilities that disrupt the smooth spherical symmetry, causing the
surface to be corrugated, and leading to variations in the shell
density.

The parameters of the bubble that are important towards our
investigation are the radius of the bubble and the swept-up mass.  The
theory of wind-blown bubbles was first derived by
\citet{weaveretal77}. The radius of the bubble can be written as:

\be 
R_b = 0.76 \left[\frac{L_w}{{\rho}_a}\right]^{1/5} \; t^{3/5} \;\;
{\rm cm}
\label{eq:rbub}
\ee

\noindent
where $L_w = 0.5 \dot{M} v_w^2$, which has dimensions of energy over
time, is called the mechanical wind luminosity, ${\rho}_a$ is the
ambient density and $t$ is the age, all in cgs units.  The massive
stars that we are considering here have lifetimes of 3.7-5 million
years depending on their initial mass. The lifetime of these stars is
a complicated function of mass and metallicity, because the mass-loss
from the star, which strongly affects its lifetime, is a function of
the metallicity. \citet{schaerer98} gives the lifetime of solar
metallicity stars as a function of mass as

\be
{ \log}\; {\tau}_{total} = 9.986 - 3.496\; \log(M) + 0.8942\; {\log(M)}^2
\ee

\noindent
where ${\tau}_{total}$ is the lifetime in years and $M$ is the star's
initial mass in solar mass units. For a 40 $\msun$ star this gives 4.8
million years.

The wind mass-loss rates and wind velocities vary over the stellar
types, and continually over the evolution, and are an even more
complicated function of the mass, luminosity, and Eddington parameter
of the star. However, a representative mass-loss rate of 10$^{-6}
\,\msun$ yr$^{-1}$ and wind velocity of 1000 km s$^{-1}$ in the main
sequence phase gives a value of $L_w > 10^{35}$ g cm$^2$ s$^{-3}$ in
the main sequence phase, increasing perhaps up to $ 10^{36}$ g cm$^2$
s$^{-3}$ in the W-R phase.

The swept-up mass depends on how far the bubble shell can expand, and
the surrounding density.  Wind-driven nebulae around W-R stars are
found to be around 3-40 pc in size \citep{cappaetal06}. The
surrounding ISM density is usually around a few, rarely exceeding 10
cm$^{-3}$ \citep{cappaetal03}. If we assume a value of 10 cm$^{-3}$ in
equation \ref{eq:rbub} and the value of $L_w$ appropriate for the
main-sequence phase (since that is where the star spends 90\% of its
life), we get a radius $R_b$ = 27.5 pc for a 40 $\msun$ star for
example. Lower densities will give larger radii. Furthermore, if the
surrounding pressure is high, the bubble pressure comes into pressure
equilibrium with the surroundings in less than the stellar lifetime,
after which the bubble stalls.

The mass of the dense shell is the mass swept up by the bubble shock
front up to that radius, and therefore the mass of molecular cloud
material up to that radius. Values of the density and volume of the
thin shell, and therefore the swept-up mass, are difficult quantities
to infer both theoretically and observationally. From those that have
been estimated, we find that observed swept-up shell masses have a
maximum value of around 1000 $\msun$ \citep{cappaetal03}, with a small
number exceeding this value. The swept-up shell mass in our
calculations is therefore taken to be 1000 $\msun$.

\subsection{\altwentysix ~from Massive Stars compared to the Early solar System}
\label{sec:al}

How do the \altwentysix amounts emitted by massive stars (\S
\ref{sec:alyields}) compare with the amount of \altwentysix required
to explain the observed value in meteorites?  The initial
concentration of \altwentysix at the time of formation of the solar
system can be calculated taking the recommended abundances for the
proto-solar system from \citet{lodders03}, which gives N(H)=2.431
$\times 10^{10}$, N(He)=2.343 $\times 10^{9}$, and
N(\altwentyseven)=8.41 $\times 10^{4}$. Using the ratio of
\altwentysix/\altwentyseven = 5 $\times 10^{-5}$ \citep{lpw76}, we can
write the mass of \altwentysix per unit solar mass (the
`concentration' of \altwentysixns) at solar system formation as
 
\be 
\label{eq:alconc}
C_{\altwentysix,pss}= \frac{5 \times 10^{-5} \times 8.41 \times 10^4
  \times 26}{2.431 \times 10^{10} + 2.343 \times 10^9 \times 4} = 3.25 \times 10^{-9} 
\ee

\noindent
This is consistent with the value given in \citet{gm12}.

We assume that the solar system is formed by the collapse of dense
material within the dense shell swept-up by the W-R bubble. Some
fraction $\eta$ of the \altwentysix produced is mixed throughout the
dense shell, giving a resulting concentration that is equal to or
greater than that computed above. Since \altwentysix is not produced
in the W-R stage, but is mixed in with the shell during this stage, we
allow for the decay of the \altwentysix in the W-R phase, which lasts
for a maximum period $t_d=$ 300,000 years (we do not consider smaller
W-R periods because given the half-life of $\sim$0.7 Myr, a smaller
W-R phase will not lead to significant decay and thus further bolster
our arguments). We use a half-life of \altwentysix of 7.16 $\times
10^5$ years \citep{rightmireetal59, samworthetal72}. After
\altwentysix is all mixed in, we assume that some fraction of the
shell collapses to form a dense molecular core that will give rise to
the proto-solar nebula.

The \altwentysix concentration after mixing with the dense shell is
given as:

\be
C_{\altwentysix,bub}= \frac{\eta M_{\altwentysix}}{M_{shell}} e^{-t_d\, ln(2)/t_{1/2}}
\ee

\begin{figure}[htbp]
\includegraphics[width=\columnwidth,angle=90]{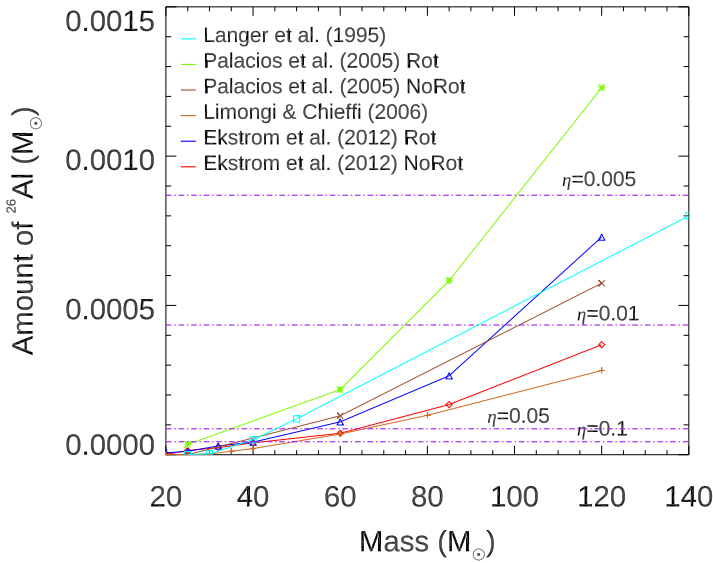}
\caption{The amount of \altwentysix emitted is taken from Figure
  \ref{fig:yields}. The dashed purple lines indicate the minimum mass
  of \altwentysix required for different values of the fraction
  $\eta$, to give the initial solar system concentration,. Where these
  thresholds cross the curves for the various \altwentysix amounts
  gives the minimum mass star needed to fulfill the requirements.}
\label{fig:alconc}
\end{figure}

The swept-up shell mass in our calculations has been assumed to be
1000 $\msun$ as mentioned above, which is on the upper end of observed
shell masses. The concentration will be inversely proportional to the
shell mass, so it can always be scaled to different masses. Lower
masses are not a problem, as this just makes it easier to achieve the
required concentration. Higher shell masses would make it more
difficult for the required \altwentysix concentration to be achieved,
but observations show a much higher mass would be quite unusual.

Figures \ref{fig:alconc} shows the amount of \altwentysix produced,
and the values of $\eta$ that satisfy the equality
$C_{\altwentysix,bub} = C_{\altwentysix,pss}$, for a decay period of
300,000 years. Since all the \altwentysix is not expected to mix with
the dense shell, a maximum value of $\eta = 0.1$ is assumed, as well
as a minimum value $\eta = 0.005$ below which no reasonable solution
is found. It can be seen that a range of solutions exists, depending
on which stellar models and values of \altwentysix production one
considers. If one assumes an efficiency of 10\% to be reasonable, then
stars above around 50 $\msun$ could produce the required \altwentysix
concentration. Lower efficiencies require higher mass stars as
expected; an efficiency of 1\% can only be matched for some models
with stars of mass $> 100 \msun$, and cannot be achieved at all in
most models. The newer stellar models produce less $^{26}{\rm Al}$,
and thus require higher efficiencies. The efficiency also depends on
the mass of the shell. If the shell mass is higher, the efficiency
will be correspondingly lower for a given \altwentysix mass.  Most W-R
shell masses though do not exceed a thousand $\msun$
\citep{cappaetal03}.

\begin{figure}[htbp]
\includegraphics[width=\columnwidth,angle=90]{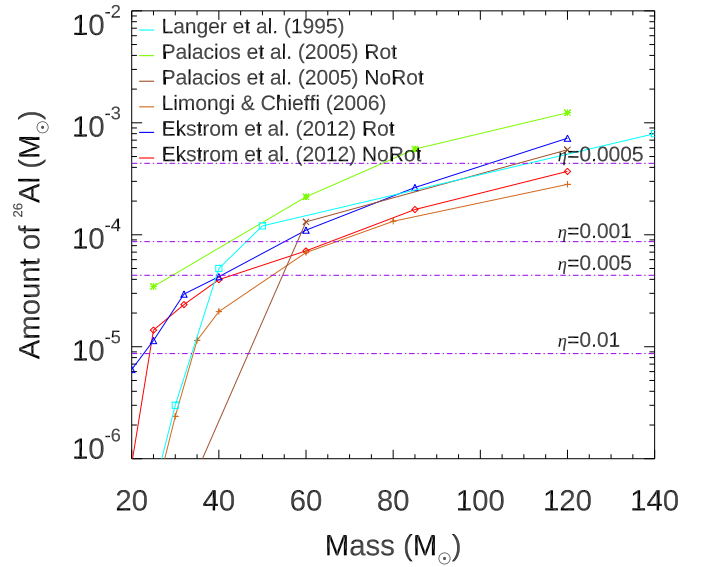}
\caption{Similar to Figure \ref{fig:alconc}, but for a minimum solar
  mass nebula. Now it is clear that even demanding a low 1\%
  efficiency, most models suggest that all W-R stars would be able to
  satisfy the initial \altwentysix requirements.  }
\label{fig:alconcmin}
\end{figure}

Another possibility exists which considerably eases the issue of
adequate \altwentysix concentration in the early solar system. While
the \altwentysix concentration is determined for the meteorites, it is
uncertain whether the same amount was present in the Sun. Indeed, some
early formed refractory condensates found in meteorites and known as
FUN (with fractionation and unidentified nuclear isotope anomalies)
calcium-aluminium inclusions, lack $^{26}{\rm Al}$, raising the
possibility that \altwentysix was heterogeneously distributed in the
early solar system and thus the abundance of \altwentysix inferred
from non-FUN CAIs may not be relevant to the solar system as a
whole. It could be that this concentration is representative of the
protoplanetary disk around the Sun but not the star itself. One could
thus assume that the \altwentysix is only present in a minimum mass
solar nebula \citep{weidenschilling77}, with mass about 0.01
$\msun$. In that case, equation \ref{eq:alconc} would be multiplied by
0.01, thus reducing the concentration to $C_{\altwentysix,min} = 3.25
\times 10^{-11}$.  This would make it much simpler for the
requirements to be satisfied. Even assuming only 1\% of \altwentysix
is injected into the dense shell, we find that in many scenarios,
almost all stars which make \altwentysixns, above about 25 $\msun$
depending on which models are adopted, would be enough to satisfy the
\altwentysix requirement (Figure \ref{fig:alconcmin}).

\section{Star formation Within Wolf-Rayet Bubbles}
\label{sec:trigstar}
Observations of molecular and infrared emission from dense neutral
clouds adjacent to OB associations led \citet{el77} to postulate that
ionization fronts, and the associated shock fronts, trigger star
formation. This was especially seen to occur near the interface
between HII regions (regions of ionized H) and dense clouds, where the
ionization front is expected to be located. Massive stars formed in
this manner then evolve and give rise to HII regions surrounding them,
which leads to another generation of star formation. Thus the process
could lead to sequential star formation. This is generally now
referred to as the ``collect and collapse'' mechanism for star
formation, because the ionization front collects the material between
it and the shock front, leading to an increase in density, which
causes the cloud material to collapse and form stars. These conditions
are exactly the ones prevalent at the edge of wind-blown bubbles,
where the ionization front is being driven into the dense shell,
leading to formation of cloud cores and a new generation of stars. The
timescale for triggering is proportional to $(G \rho_{sh})^{-1/2}$,
where $G$ is the Gravitational constant and $\rho_{sh}$ is the density
of the collapsing material (dense shell). Generally, the most unstable
wavelength is comparable to the shell thickness, as would be expected
for thin shell instabilities. Analytic studies of the process were
completed by \citet{whitworthetal94a,whitworthetal94b}. Simulations
\citep{hi05, hi06a,hi06b} have shown that shells driven into molecular
clouds do have time to collapse and form stars, and that triggered
star formation does work.

Another somewhat similar mechanism where the ionization front ablates
the cloud, generating a shock that compresses the cloud, causing
clumps to collapse, is the radiation-driven implosion model,
elaborated on by \citet{ll94} as a way to explain bright rimmed clouds
and cometary globules. This operates on a shorter timescale compared
to the previous mechanism, and also in a smaller spatial region. The
actual mechanism is hard to distinguish and not always clear.
\citet{walchetal15} suggest that a hybrid mechanism, that combines
elements of both processes, may be at work in the HII region RCW120.
Recent reviews of both observational and theoretical aspects of
triggered star formation can be found in \citet{elmegreen11a} and
\citet{walch14}.

Observational evidence for triggered star formation has been found at
the boundaries of wind-bubbles around massive O and B stars
\citep{deharvengetal03,dzc05,zavagnoetal07, watsonetal09,
  brandetal11}. Further evidence arises from statistical correlation
of young stellar objects with wind bubbles. \citet{kendrewetal12}
investigated infrared bubbles in the Milky Way project and found that
two-thirds of massive young stellar objects were located within a
projected distance of 2 bubble radii, and about 1 in 5 found between
0.8 and 1.6 bubble radii. Furthermore, as the bubble radius increased
(and thus the swept-up shell mass increases) they found a
statistically significant increase in the overdensity of massive young
sources in the region of the bubble rim.  Also pertinent to the
current discussion, molecular cores undergoing gravitational collapse
due to external pressure from the surrounding gas have been found
around W-R star HD 211853 \citep{liuetal12}. The cores have been
estimated as being from 100,000 to greater than a million years
old. Thus, both observational and theoretical considerations suggest a
high probability of triggered star formation at the boundaries of
wind-blown bubbles, where suitable conditions are both predicted and
observed.

\section{Transport of \altwentysix}
Another important ingredient needed to evaluate the plausibility of
the W-R injection scenario is the mechanism for injection of
\altwentysix from the wind into the dense shell.  One aspect that has
been conspicuously absent is a discussion of the mechanism of mixing
of the \altwentysix from the hot wind into the cold dense
shell. \citet{gm12} attributed it to turbulent mixing without giving
further details. \citet{young14, young16} assumes mixing from winds to
clouds without providing details of the process. \citet{gaidosetal09}
suggested that dust grains were responsible for the delivery of
\altwentysix from W-R winds into the molecular cloud, where they were
stopped by the high density clouds. However they did not investigate
the properties of the dust grains seen around W-R stars. \citet{tdd10}
realized that mixing of the hot winds into the colder and denser
material was a difficult problem, but refuted the arguments of
\citet{gaidosetal09} on the basis that (1) grains (assumed to be about
0.01 $\mu$m) would be stopped before they reached the dense shell and
(2) that the mean velocity for the emitting \altwentysix nuclei of
about 150 \kms, derived from the broadening of the 1.8 MeV line seen
with RHESSI and INTEGRAL gamma-ray satellites, was too low for them to
have survived sputtering at the reverse shock. Instead they formulated
a model where the mixing is due to instabilities in the bow shock
region created by a runaway W-R star that moves relative to the center
of mass of the W-R bubble. These instabilities tend to mix the W-R
wind material with the surrounding material, which is most likely
material ejected during the star's prior evolution, mixed in perhaps
with some pre-existing material.

The mixing of fast hot material (such as from the winds) with slower
cold material (as in the dense shell), has been more thoroughly
studied in the context of injection by a SN shock wave, and insight
can be gained from those results.  \citet{boss06} and \citet{bk13}
have shown that the injection efficiency due to hydrodynamic mixing
between a SN shock wave and the collapsing cores is small, of order a
few percent. In their model the mixing occurs late in the SN
evolution, when it has reached the radiative stage and the SN shock
has slowed down to $<$ 100 km s$^{-1}$. This requires the SN to be
several pc away from the initial solar system so that the shock can
slow down from its initial velocity of thousands of km s$^{-1}$. The
SPH simulations of \citet{gv00} also show that shock velocities
between 20 and 45 \kms are required to trigger collapse. However their
calculations with a variable adiabatic exponent $\gamma$ appear to
suggest that the hot shocked gas is still unable to penetrate the cold
cores due to buoyancy and entropy effects, further complicating the
issue. An alternate model by \citet{odh07} proposed a SN that was much
closer (0.3 pc) to the proto-solar nebula. This model though did not
account for the ionizing radiation from the progenitor star
\citep{tdd10}. \citet{mvl84} showed that massive stars that
core-collapse as SNe could clear out large regions of space around
them due to their ionizing radiation. Thus the disk would be adversely
affected even before the star collapsed to form a SN, and it is not
clear how viable this model is.

The W-R wind velocity substantially exceeds the SN shock velocity
discussed above of 100 \kms. W-R winds also have a much lower density
than SN ejecta, since the density is proportional to the mass-loss
rate and inversely to the velocity.  The efficiency of mixing in winds
will therefore be further reduced.  Furthermore, winds sweeping past
high-density cores will lead to shearing at the edges of the cores,
leading to the growth of Kelvin-Helmholtz instabilities at the
interface, and essentially stripping material away. Some material will
be mixed in, but it will be a very small fraction. Hydrodynamic mixing
does not appear a viable mechanism in this scenario. We suggest
instead that \altwentysix condenses onto grains that serve as
injection vectors into the bubble shell, a conclusion that was also
reached by \citet{gaidosetal09} for stellar wind injection and
\citet{odh10} for SN injection.

\subsection{Dust in Wolf-Rayet Bubbles}

Infrared emission, indicative of dust, has been seen in carbon-rich
W-R (WC) stars since 1972 \citep{ash72}.  Even early on, it was
realized that circumstellar dust emission was present mostly around
the late carbon-rich sub-types of W-R stars, WC8-WC10 stars
\citep{wvt87}, as well as a WN10 (N-rich W-R) star. The clearest
manifestation of dust however was the observation of a `pinwheel
nebula' around WR 140 \citep{tmd99}. In this case it was clear that
the presence of interacting winds due to a companion star was
responsible for the formation of dust. This was followed by the
discovery of another pinwheel nebula around the star WR 98a, further
solidifying the binary connection. It is conjectured, although not
conclusively shown, that only WC stars out of all the W-R stars are
capable of producing dust, either steadily or episodically
\citep{williams02}. Even though the number of stars is small, dust
seen in WC stars is important because the absolute rate of dust
production is found to be high ($\sim 10^{-6}\, \msun yr^{-1}$ of
dust) \citep{mm07}, and a few percent of the total wind mass. Analysis
of the IR emission showed that dust forms close in to the star.

Modelling the IR emission from WR 112, \citet{marchenkoetal02}
suggested a grain size of 0.49 $\pm$ 0.11 $\mu$m. This is consistent
with the results from \citet{ct01} who inferred grain sizes of around
1 $\mu$m from ISO spectroscopy and analysis of the dust around the
stars WR 118, WR 112 and WR 104.  Dust grain sizes of around
0.3-2$\mu$m are also inferred from modelling the dust around WR 95 and
WR 106 \citep{rajagopaletal07}. Thus it appears that dust grains
formed in WR bubbles have sizes predominantly around 1 $\mu$m. It has
been now shown that dust can be formed around WC stars, Luminous Blue
Variable (LBV) stars, and possibly WN stars \citep{rajagopaletal07},
although the latter is questionable. There is, however, no doubt that
dust can be formed around carbon-rich WC stars, and LBV stars which
may transition to W-R at a later stage.

\subsubsection{Condensation of \altwentysix into Dust Grains}
Given the presence of dust in WC stars, we address further the
question of \altwentysix transport. One of the arguments made by
\citet{tdd10}, that grains would not be able to survive, is negated by
the presence of mainly large $\mu$m size grains in W-R winds, which we
show later do manage to survive. Sputtering at the reverse shock does
not appear to be a problem in the low density wind for the large
grains.  Regarding the low expansion velocity indicated by the line
broadening, this is indicative of the bulk velocity of the grains and
unrelated to the issue of grain destruction. As we show later, in our
model \altwentysix takes only about 20,000 years to travel from the
star to the dense shell, after which its velocity would decrease
considerably before coming to a complete stop. This is much smaller
than the lifetime of 0.7 Myr, thus at any given time we would expect
most of the \altwentysix to be expanding at a low velocity in the
dense shell rather than at the wind velocities. The Doppler velocity
and therefore linewidth would be representative of the average bulk
velocity, and therefore be dominated by the low velocity \altwentysix
decaying in the cavity walls. In fact such a scenario was already
envisioned by \citet{diehletal10}, who explained the low expansion
velocity by postulating that \altwentysix was being slowed down, on a
scale of about 10pc, by interacting with the walls of pre-existing
wind-blown cavities in the region. Our work shows how this scenario
would work in practice.

One important question is whether \altwentysix would condense into
dust grains. Equilibrium calculations \citep{fmg10} suggest that would
be the case. In general, in stars such as LBVs, the outer layers would
condense to give aluminium oxides. In the carbon-rich WC stars, AlN
would be the dominant Al condensing phase. One must also not neglect
the fact that most dust-producing WC stars are in a binary system, and
the companion is generally an O-type MS star, thus suggesting that H
may still be present in the system, leading to the formation of
hydrocarbons. In the final, oxygen-rich phase of W-R stars (WO), it
would be the oxides that would dominate again. Although the
calculations by \citet{fmg10} are equilibrium calculations, and
consider only stars up to 25 $\msun$, they do suggest that most of the
\altwentysix would condense into dust grains. \citet{odh10} also
discuss this issue in detail, providing arguments drawn from theory
and astronomical observations to support the idea that the
condensation efficiency of \altwentysix is around 10\%.

\subsubsection{Injection of \altwentysix into the Solar nebula}
Once the \altwentysix condenses into dust grains, this dust must
travel several pc from the vicinity of the star out to the dense
shell. Grains can be destroyed by thermal sputtering at the reverse
shock of the wind blown bubble. According to \citet{ds79}, the
lifetime against thermal sputtering in the hot gas is:

\be
t_{sput} \approx 2 \times 10^4 \, \left[\frac{n_H}{{\rm cm}^{-3}}\right]^{-1}\, \left[\frac{a}{0.01 \mu {\rm m}}\right]\; yr
\label{eq:sputter}
\ee
\noindent
where $n_H$ is the number density within the bubble, and $a$ is the
size of the dust grain. For typical wind bubbles evolving in the
interstellar medium, bubble densities are of order $n_H = 10^{-2}$
cm$^{-3}$. Using a grain size of $a=1{\rm \mu}$m, the lifetime is
10$^8$ years, almost two orders of magnitude greater than the stellar
lifetime.  The interior density of the bubble goes as
\citep{weaveretal77, vvd07} ${\rho}_{am}^{3/5}$, so that even if the
external density were as high as 10$^3$ cm$^{-3}$ (which is highly
unlikely over the entire bubble expansion), the bubble internal
density is about 0.6, and the lifetime against sputtering still
exceeds a million years. This could be a sizable fraction of the
star's lifetime, and certainly exceeds the duration of the W-R phase
of any massive star. Furthermore, calculations of C dust destruction
in SNe have shown that C grains are the most resistant amongst all
species to sputtering \citep{nozawaetal06, bc16} and are therefore
most likely to survive sputtering at shocks. Thus over the entire
relevant parameter range, thermal sputtering can be considered
negligible for the large size dust grains found in W-R bubbles.

The stopping distance ($d$) of a grain with size $a$ and grain density
${\rho}_g$ in the interstellar bubble with density ${\rho}_b$ is given
by $d = {\rho}_g/{\rho}_b \times a$ \citep{spitzer78}. For $\mu$m size
grains with grain density 2~g cm$^{-3}$ in a bubble with internal
density $n_H = 10^{-2}$ cm$^{-3}$, the stopping distance is found to
be of order 3000 pc, far larger than the size of the bubble in the
high density molecular cloud. Even for a large interior density of
$n_H = 1$ cm$^{-3}$, which is highly unlikely, the stopping distance
of 30pc is comparable to the size of the bubble.  Even if we take into
account the fact that this formula may overestimate the range
\citep{ragot02}, it is still clear that for most reasonable parameters
the dust grains will survive passage through the bubble. Indeed,
\citet{marchenkoetal02} find observationally in the case of WR 112
that about 20\% of the grains may be able to reach the interstellar
medium.

The dust grains are carried by the W-R wind out to the dense
shell. Once the wind reaches the dense shell, it is strongly
decelerated, but the grains will detach from the wind and continue
moving forward into the dense shell. The stopping distance of the
grains will now be reduced. Since the shell density is 10$^4 - 10^5$
times higher than the density in the bubble interior, the stopping
distance correspondingly reduces from pc to 10's to 100's of
AU. Following \citet{odh07}, the time taken for the dust to lose half
its initial velocity is $t_{1/2} = 16 {\rho}_g\, a/(9\,{\rho}_b \,
v_w)$, which is only 35 years using the shell parameters. It is clear
that the grains will be stopped in only a few hundred years.

Thermal sputtering time of grains with size $a$ can be written as

\be
t_{sput} = a\left[\frac{da}{dt}\right]^{-1} \;\;\; {\rm s}
\ee

The rate at which the radius of the dust grain changes ($da/dt$) can
be written as a function of the plasma temperature $T$ and number
density $n_H$ \citep{tm95}:

\be \frac{da}{dt} = -3.2 \times 10^{-18}\; n_H \, \left[\left(\frac{2
    \times 10^6}{T}\right)^{2.5} + 1\right]^{-1} \;\;\; {\rm cm \;
  s^{-1}} 
\ee

Note that for high temperatures $> 2 \times 10^6$ K characteristic of
the bubble interior, this reduces to equation
\ref{eq:sputter}. However, the dense shell will have much lower
temperatures, on order $10^4$ K, with denser regions having even lower
temperatures. The sputtering time at these low temperatures increases
considerably:

\be
t_{sput} = 5.6 \times 10^{11} \, \left[\frac{n_H}{{\rm cm}^{-3}}\right]^{-1}\, \left[\frac{a}{1\, \mu {\rm m}}\right]\;\left[\frac{T}{10^4 \,{\rm K}}\right]^{-2.5} yr
\ee

\noindent 

The density of the dense shell is higher than the density of the
swept-up material around it, and thus could be 10$^3$ to 10$^4$
cm$^{-3}$.  Even the high density would give a lifetime against
sputtering of order several tens of millions of years, so thermal
sputtering becomes irrelevant at this stage due to the low
temperatures.

However other processes, such as grain evaporation, and non-thermal
sputtering due to high speed collisions between the dust grains and
the dense gas, become important during this phase. The relative
velocity of the fast grains impacting the dense shell can exceed
several hundred km s$^{-1}$. At these velocities, the grains will
experience frictional heating which will raise their temperatures to
greater than dust condensation temperatures \citep{odh10}, which are
in the range of 1300-1700 K \citep{lodders03}. Thus the impact can
cause the grains to vaporize, releasing the \altwentysix into the
dense gas. Non-thermal sputtering, which is independent of the
temperature but depends on the relative velocity between the grains
and the shell gas, also becomes important at this stage. The rate of
non-thermal sputtering is close to its maximum value, and comparable
to the value of thermal sputtering, at a relative velocity of about
1000 $\kms$ \citep{goodsonetal16}, which is approximately the value at
which the grains impact the dense gas in the shell. Thus at the
densities within the shell, the lifetime against sputtering at impact
will be on order 1000-10000 years. A fraction of the grains will be
sputtered away. As the velocity decreases, this lifetime will
increase. Due to frictional drag within the disk \citep{odh10} the
grains will increase in temperature and some will evaporate.

Estimating what fraction of the grains eventually survive requires
numerical simulations taking all the various processes into account,
which is beyond the scope of this paper. Simulations under somewhat
similar conditions were carried out by \citet{odh10} and
\citet{goodsonetal16} for dust grains impacting a dense disk. While
the impact velocities and other details differ, the results clearly
show that a large fraction of grains 1 $\mu$m in size will penetrate
the dense disk and inject between 40\% \citep{goodsonetal16} to 80\%
\citep{odh10} of the SLRs into the dense shell. This is contrasted
with direct injection of SLRs by the gas, which was found to be only
1\% by \citet{goodsonetal16}, comparable to the previous results of
\citet{bk13}. \citet{odh10} suggest that up to 40\% of the 1 $\mu$m
grains, and about 30\% of grains of all sizes, may survive within the
dense disk, although they have not carried their simulations on for
much longer to determine whether they will survive or be eventually
sputtered or evaporated away. \citet{goodsonetal16} estimate that
about half of all grains sputter or stop within the cloud, but the
velocity of their shock on impact, $\sim$ 350 $\kms$, is smaller than
the velocity that the grains in our simulation would have. A larger
velocity would increase the amount of non-thermal sputtering.

Using these results as well as our own estimates above, one may expect
at least half the \altwentysix that reaches the dense shell to be
injected from the grains to the disk. Even if one assumes that about
20\% of the \altwentysix condenses into dust grains, and half the
grains survive within the bubble interior and reach the dense shell,
both conservative estimates, we find that a total of $0.2 \times 0.5
\times 0.5 = 0.05$ of the \altwentysix from the star will reach the
dense shell. Thus efficiencies of this order used to calculate the
\altwentysix fraction in section \ref{sec:al} were warranted.
Furthermore, it appears from the simulations, and assuming the grains
are distributed in size around the 1 $\mu$m value, that at most 1/3 of
the grains may eventually survive.

We investigate further if the survival of these grains is consistent
with observations of meteorites. Our simulations show that there is
extensive mixing in the bubble, so although the \altwentysix and C may
be emitted at somewhat different times, the material can be assumed to
be completely mixed in the W-R stage. We therefore consider the bulk
composition of the material to determine what the composition of the
grains might be. In the models of \citet{ekstrometal12}, the mass
ratio of \altwentysixns/C in non-rotating stars between initial mass
32 and 120 $\msun$ (the likely mass to form W-R stars) varies between
10$^{-3}$ to 7. $\times 10^{-5}$, with stars above 60 $\msun$ having
ratios generally a few times 10$^{-5}$. For rotating stars this ratio
lies between 2.6 $\times 10^{-5}$ to 3.$\times 10^{-4}$. For
simplicity in this calculation we consider a mass ratio 5$\times
10^{-5}$.

The mass fraction of \altwentysix in the initial solar system is given
by equation \ref{eq:alconc} to be 3.25 $\times 10^{-9}$. If we assume
that all the \altwentysix comes from the W-R star, the fraction of C
that then arises from the W-R star, with respect to H and He, is \be
\frac{3.25 \times 10^{-9}}{5 \times 10^{-5}} = 6.5 \times 10^{-5}.
\ee

What we are really interested in is the mass fraction of C relative to
condensable matter found in meteorites, rather than H and He. For the
proto-sun, the fraction of metals (essentially everything but H and
He) is 0.0149 \citep{lodders03}. The main contributors to this are C,
N, O, Si, Mg and Fe, with others contributing to a lesser
degree. Subtracting species that do not condense such as the rare
gases, we find that the mass-fraction of condensable matter is about
1.3 $\times 10^{-2}$. Therefore the concentration of C arising from
W-R grains, relative to condensable matter, is 5 $\times 10^{-3}$. If
one-third of these grains survived injection into the dense shell and
proto-solar system, then the mass-fraction of C grains in the
proto-solar system, and thereby in meteorites, would be about
0.16\%. This is a large and potentially identifiable fraction.

However, we must also consider that there are processes which destroy
dust in the early solar system compared to the dust concentrations in
the neighbourhood. \citet{zhukovskaetal08} have studied the evolution
of various dust species in the ISM at the time of solar system
formation. They predict that the mass-fraction of C dust compared to H
is about 1500 parts per million (ppm) at time of solar system
formation. If one considers dust produced by SNe and AGB stars only,
it is a factor of 10 less or about 150 ppm. On the other hand, the
mass-fraction of C-containing grains such as graphite and SiC grains
found in meteorites is close to 10 ppm \citep{nittler03}. Normalized
to H this would be of order 0.1 ppm. This means that there is a
destruction factor of order $\sim$1000 when going from interstellar
dust to dust grains in meteorites. If the same factor was applicable
to the W-R grains, this would imply that only 10$^{-4}$ \% of the
grains would be potentially identifiable in meteorites, a negligible
fraction. Thus our assumption of about one-third of the grains
surviving does not appear to be a problem from the point of view of
meteoritic observations.

The dense shell is clearly not a spherically symmetric, homogeneous
shell. It is susceptible to several dynamical and radiative
instabilities, such as Vishniac-type thin-shell instabilities, or
ionization front instabilities,
\citep{dcb96,glm96,db98,vvd07,vk12,ta11,dr13} that tend to break up
the shell and wrinkle the surface, as seen in Figure
\ref{fig:wrbub}. The density within the shell is also not completely
uniform, since it depends on the surrounding density which may vary
over the circumference, the penetration of the ionization front into
the shell, and the disruption due to various instabilities. The result
is that both the inner radius and density are somewhat variable.  The
dust grains themselves, although having an average size of 1 $\mu$m,
will have a range of sizes around that value. Thus the penetration
depth and stopping distance of the dust grains into the dense shell
will vary along the circumference of the shell, introducing a level of
heterogeneity into the picture of \altwentysix injection into the
collapsing cores.

\section{The Subsequent Supernova explosion}
At the end of the W-R stage, the star will end its life in a
core-collapse followed by a stellar explosion, giving rise to a
supernova shock wave and leaving a compact remnant behind. Some
massive stars are predicted to core-collapse all the way to a
black-hole, leaving no remnant, and the dividing line is not clearly
delineated. In the work of \citet{georgyetal12}, stars all the way up
to masses of 120 $\msun$ can form a Type Ib/c SN if the formation of a
black hole during the process has no influence on the resulting SN
explosion. Conversely, if the formation of a black hole does not
result in a bright SN explosion, then only stars up to about 34
$\msun$ (44 $\msun$ in the non-rotating limit) will form Type Ib/c
SNe. \citet{sw16} emphasize that the explosion properties are
sensitive to the mass-loss prescription employed. Whether there will
be a faint SN, or no SN at all, depends on whether the star can
successfully launch an outward moving shock wave (not always the
case), and whether there is fallback of the $^{56}{\rm Ni}$ into the
center \citep{whw02, sw16}. Thus, while it is realistic to assume that
W-R stars will form Type Ib/c SNe, up to what initial mass that
happens is unclear.

In this work, we have found that we would need a massive star to seed
the \altwentysix in the initial solar system. Depending on the
efficiency of mixing \altwentysix with the surrounding dense shell, in
some scenarios we would need a star $> 40 \; \msun$. The star either
ends its life in a spectacular supernova explosion, or it directly
core-collapses into a black hole \citep{whw02}. If there is no SN
explosion then there is no explosive nucleosynthesis, and no resulting
shock wave. This will mean that there is no \fesixty produced during
the explosion. There is still some \fesixty produced in shell C
burning (in stars $< 60\, \msun$) and in shell He burning (in stars $>
60\, \msun$) \citep{lc06}, which could be ejected, but this material
has low velocity and is not pushed by the shock wave into the dense
shell. The fraction mixed in with the shell would be considerably
reduced from the few percent expected from earlier calculations, to an
extremely tiny fraction. Overall we would not expect any significant
amount of \fesixty from the fallback SN.

If there is a SN explosion, then explosive nucleosynthesis will take
place, accompanied by a shock wave, and there will be production and
ejection of \fesixty.  In the simulations of \citet{boss06, fb97,
  bk13}, it is the transmitted shock into the dense disk, and the
associated Rayleigh-Taylor instabilities, that injects SN ejecta into
the disk. Therefore this depends crucially on where and how far behind
the \fesixty is located in the ejecta. The amount of \fesixty that
might be injected to the solar system may be estimated by considering
the example of the 40 $\msun$ stellar model in
\citet{vanmarleetal05}. By the time the star ends its life, 31.8
$\msun$ of material is lost via mass-loss to the surrounding medium,
which mass is enclosed within the bubble. If 1.5 $\msun$ is assumed to
be left behind in a neutron star, the ejected mass will be 6.7
$\msun$. 6.7 $\msun$ sweeping up 31.8 $\msun$ will mean the SN will
not have reached the Sedov stage \citep{dc98}, which requires the
swept-up mass to be much higher.  The expanding shock front will have
a double-shocked structure consisting of a forward and reverse shock,
separated by a contact discontinuity. Iron-60 can penetrate the dense
shell/core if it has already been shocked by the reverse shock and
lies near the contact discontinuity which is unstable.  For this to
happen, the \fesixty must be in the outermost, higher velocity layers
which are initially shocked by the reverse shock. Iron-60 is formed in
the He or C convective shells, although in stars above 40 $\msun$ the
major contribution is from the He burning convective shell
\citep{lc06}.  One would expect the W-R star to have shed its H, and
presumably some portion of its He layer.  If \fesixty is mainly formed
in the He-shell, then it will lie quite close to the outer edge of the
ejecta, and the reverse shock will likely reach it before the forward
shock collides with the dense shell.  Given that the contact
discontinuity is always Rayleigh-Taylor unstable \citep{cbe92,vvd00}
it is possible that parts of the shocked ejecta forming the unstable
Rayleigh-Taylor `fingers' that penetrate into the shocked ambient
medium, may come into contact with the dense shell/cores. Even then, a
further complication is that if the SN forward shock wave has speeds
exceeding 1000 \kms, the post-shock gas will be at temperatures $>
10^7$K, and will have a difficult time penetrating the colder
material, as pointed out earlier \citep{gv00}.

The above arguments assume that the \fesixty is uniformly deposited
and the shock wave is spherically symmetric. This is not necessarily
the case. W-R stars are also thought to be the progenitors of
gamma-ray bursts, where the emission is highly beamed in a jet-like
explosion, and thus is highly asymmetric. Although these are the
extreme cases, it has been shown, especially from observations of
double-peaked profiles in the nebular lines of neutral oxygen and
magnesium, that explosions of W-R stars, which lead to Type Ib/c SNe,
are generally aspherical \citep{mazzalietal05}. Some results suggest
that all supernova explosions from stripped envelope stars have a
moderate degree of asphericity \citep{maedaetal08}, with the highest
asphericities occurring for SNe linked to gamma-ray bursts. In the
current scenario, we can assume that given the small size of the
proto-solar nebula compared to the dense shell, even a moderate degree
of asphericity such as a factor of 2 would result in a 50\%
probability that the SN debris, including the \fesixty, would not
reach the fledgling solar system at all.

Therefore we conjecture that there is a high degree of probability
that after the death of the W-R star in a core-collapse explosion, the
resulting SN ejecta containing \fesixty would not be able to
contaminate the proto-solar nebula and raise the level of \fesixty
beyond the level of the material in the swept-up dense shell.

The \fesixty within the solar nebula in our model arises in the
swept-up material that forms the dense shell. In a steady-state
situation, the abundance of \fesixty in the swept-up gas is a result
of the Galactic evolution up to the beginning of the solar system,
including many past generation of stars. Assuming that the \fesixty
had reached an equilibrium situation, the gas should have an abundance
of \fesixty equal to the Galactic value as discussed in \S 1. However,
since the star's lifetime is greater than the lifetime of \fesixty by
up to a factor of 2, some of this material will radioactively
decay. On average the shell gas will therefore have an \fesixty
abundance that is slightly lower than the Galactic value.

\section{ Formation of the Solar Nebula:} 

In our model, the stellar wind reaches the dense shell only after the
onset of the W-R phase. After this wind actually reaches the shell, it
is decelerated, the dust grains detach from the wind, and are injected
into the dense disk, where they are stopped or sputtered away. This
process thus takes place in the latter part of the W-R phase, during
the WC phase when dust is formed. The lifetime of the WC phase varies
from about 1.5 $\times 10^5$ yr for a 32 $\msun$ star to about 3
$\times 10^5$ yr for a fast rotating 120 $\msun$ star
\citep{georgyetal12}. After the onset of the W-R phase, the wind still
takes some time to reach the shell, on order of a few to 10,000
years. The wind is decelerated on impact with the shell but the dust
grains detach and continue with the same velocity into the shell. The
entire process of the wind being expelled from the star and carrying
the dust with it would take less than 2 $\times 10^4$ yr. We would
expect that the timescale for \altwentysix to be injected and mixed in
with the dense shell is less than 10$^5$ years after \altwentysix
ejection from the star. Uncertainties include how long the dust takes
to form, and how long the \altwentysix takes to condense into the dust
grains. Theoretical arguments of the distribution and ages of CAIs in
the solar nebula \citep{ciesla10} suggest that the mixing of
\altwentysix with the pre-solar material took place over about 10$^5$
yr, so our results are consistent with this. During this period of
mixing, some portion of the dense shell was collapsing to form
molecular cores, including the one that gave rise to our solar system.

The timescale for the triggered star formation is shorter for the
radiation-driven implosion mechanism than the collect and collapse
mechanism. The average time for fragmentation to start in expanding
shells is of order 0.9 Myr \citep{whitworthetal94b}. Given that the
lifetimes of these stars are several million years and the W-R phase
occurs only at the end of its life, it is likely that injection of
\altwentysix happens almost simultaneously as the shell is starting to
fragment and cores begin to form.  Calculations also indicate that
heterogeneity in the initial material appears to be preserved as the
core collapses \citep[see for e.g.][]{vissieretal09}, so we would
expect that the \altwentysix distribution set up by the dust grains
will be preserved when parts of the shell collapse to form
stars. Since the \altwentysix does not penetrate all the way into the
dense shell, given the short stopping distance, it is likely that
there would be some regions which do not contain much or perhaps any
\altwentysixns.  This is consistent with the fact that FUN CAI's in
meteorites show very little to zero \altwentysix
\citep{esatetal79,armstrongetal84,macphersonetal07}. Platy hibonite
crystals, and related hibonite rich CAIs are not only \altwentysix
poor but appear to have formed with \altwentysix/\altwentyseven ratio
less than the Galactic background \citep{koopetal16}, whereas spinel
hibonite spherules are generally consistent with the canonical
\altwentysix/\altwentyseven ratio \citep{liuetal09}. In the current
scenario, such a diversity in \altwentysix abundance would be
expected. In this model our solar system is not the only one with with
a high \altwentysixns; other solar systems will be formed over the
entire disk area that may also have similar abundances of
\altwentysixns. Depending on the subsequent evolution of the star and
the formation or not of a SN, these other systems may have different
amounts of \fesixty compared to ours. Thus our prediction is that
there should be other solar systems with abundance of \altwentysix
similar to ours, but with equal or higher abundances of \fesixtyns.

\section{Discussion and Conclusions}

The inference of high abundance of \altwentysix in meteoritic material
has led to speculations over several decades that the early solar
system formed close to a SN. This however would be accompanied with an
abundance of \fesixty above the background level. The recent discovery
that the amount of \fesixty in the early solar system was lower than
the Galactic background has prompted a re-evaluation of these ideas
\citep{td12}.

\begin{figure}[htbp]
\includegraphics[width=\columnwidth]{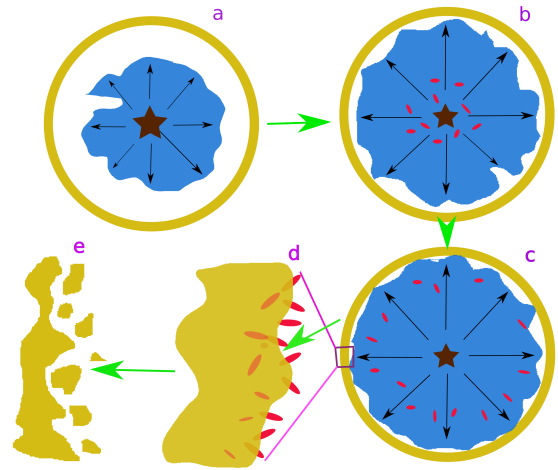}
\caption{Cartoon version of our model of the formation of the solar
  system. (a) A massive star forms, and its strong winds and ionizing
  radiation build a wind-blown bubble. The blue region is the
  wind-blown region, the yellow is the dense shell. An ionized region
  (white) separates them. (b) The bubble grows with time as the star
  evolves into the W-R phase. In this phase, the momentum of the winds
  pushes the bubble all the way to the shell. At the same time, dust
  forms in the wind close in to the star, and we assume the
  \altwentysix condenses into dust. (c) Dust is carried out by the
  wind towards the dense shell. (d) The wind is decelerated at the
  shell but the dust detaches from the wind and continues onward,
  penetrating the dense shell. (e) Triggered star formation is already
  underway in the last phase. Eventually some of the material in the
  shell collapses to form dense molecular cores which will give rise
  to various solar systems, including ours. }
\label{fig:ss1}
\end{figure}

In this work we have put forward an alternate suggestion, that our
solar system was born at the periphery of a Wolf-Rayet bubble. Our
model adds details and expands considerably on previous results by
\citet{gaidosetal09}, \citet{odh10} and \citet{gm12}, while adding
some totally new aspects, especially the fact that our solar system
was created at the periphery of a Wolf-Rayet wind bubble by triggered
star formation. W-R stars are massive stars, generally $\ge 25 \msun$,
that form the final post-main-sequence phases of high-mass O and B
stars before they core-collapse as supernovae (see Figure
\ref{fig:ss1}). The \altwentysix that is formed in earlier phases is
carried out by the stellar mass-loss in this phase. The fast
supersonic winds from these stars carve out wind-blown bubbles around
the star during their lifetime (Figure \ref{fig:ss1}a). These are
large low density cavities surrounded by a high-density shell of
swept-up material. In our model, the \altwentysix ejected in the W-R
wind (Figure \ref{fig:ss1}b) condenses onto dust grains that are
carried out by the wind without suffering much depletion, until they
reach the dense shell (Figure \ref{fig:ss1}c).  The grains penetrate
the shell to various depths depending on the shell density (Figure
\ref{fig:ss1}d). The ionization front from the star entering the shell
increases the density of the shell, causing material that exceeds a
critical mass to collapse and form star-forming cores (Figure
\ref{fig:ss1}e). This has been observed at the periphery of many wind
bubbles, and we suggest that it is what triggered the formation of our
solar system, which was enriched by the \altwentysix carried out in
the W-R wind.

In our model, the \altwentysix will be injected into the dense shell
while the shell material is collapsing to form the dense cores that
will form stars that give rise to the solar system. Thus this will
happen close to the end of the star's evolution and/or near the SN
explosion. Eventually either the SN shock wave will eat away at the
bubble and break through it; material will start to leak through the
bubble shell, which is generally unstable to various instabilities
\citep{dcb96,glm96,db98,vvd07,vk12,ta11,dr13}, until it is completely
torn apart; or the bubble shell will dissipate. Either way the nascent
solar system will then be free of its confined surroundings.

A single star is responsible for all the \altwentysix in the initial
solar system, and perhaps (see below) for most other short-lived
radionuclides (SLRs). Our model uses dust grains as a delivery
mechanism for the \altwentysix into the dense shell that will then
collapse to form the pre-solar core. In this our model differs from
many previous ones that also suggested W-R stars as the source of the
radionuclides. As pointed out, the delivery mechanism and mixing of
the SLRs with the presolar molecular cloud is of utmost importance, as
also shown by \citet{fb97}, \citet{odh10} and \citet{bk13}; it is very
difficult to get tenuous hot gas, whether carried by fast winds or
supernova ejecta, to mix with the dense, cold gas in the pre solar
molecular cloud.  \citet{gm12} considered W-R bubbles, but did not
address the crucial part of delivery and mixing of the SLRs into the
molecular cloud, attributing it to turbulent mixing. \citet{young14}
and \citet{young16} do consider stellar winds from W-R stars as well
as SNe, but did not consider the question of mixing of the wind or SN
material with the molecular cloud and the injection of SLRs, simply
assuming that it will happen by some unknown mechanism over a
relatively long time-period. \citet{gaidosetal09} did address the
mixing and also attributed it to dust, and our model is closest in
nature to theirs. However, they did not work out a complete and
detailed model as we have done here.

\citet{bk13} conclude that W-R winds are untenable as the source of
the \altwentysix due mainly to 2 reasons: (1) The high wind velocity
is not suitable for injection and (2) The wind velocities are too high
and will end up destroying the molecular cores rather than triggering
star formation. In principle we agree with both of these, and neither
of them forms an impediment to our model. We have already mentioned
that hydrodynamic mixing due to the wind is not a viable source of
mixing, hence our preferred method is via the condensation of
\altwentysix into dust grains, which survive in the low density wind
and can be injected into the dense shell.  Regarding the fact that the
winds will not trigger the star formation, star formation in our model
is not triggered by the wind, but by a mechanism that involves the
ionization and shock fronts. It should be mentioned that the forward
shock of the wind bubble is always radiative, and is observationally
and theoretically measured to move at speeds of 20-50 \kms. Perhaps
the one shortcoming of our model may be that if the molecular cores
form too early, they may still be destroyed by the wind. Nonetheless,
the eventual proof arises from observations, which clearly show
evidence of triggered star formation around numerous wind bubbles, and
especially cores undergoing gravitational collapse around a W-R star,
as mentioned in \S \ref{sec:trigstar}.

While there is significant circumstantial evidence that triggered star
formation occurs at the periphery of wind-blown bubbles, confirmation
would require that the age of the subsequently formed stars be much
smaller than that of the parent ionizing star. A further question that
remains is whether our solar system is in fact
special. \citet{whitworthetal94a} have suggested that the collapse of
the shocked layers would result in the preferential formation of high
mass fragments, although clearly this does not indicate high-mass
stars, as each fragment could easily split into several low-mass
stars. It is difficult to measure the initial mass function that
characterizes the mass-range of the newly formed stars, although some
attempts have been made. \citet{zavagnoetal06} find about a dozen
suspected massive star sources among the possibly hundreds of sources
detected in mm and infra-red images of the Galactic HII region
RCW79. \citet{deharvengetal06} observe 2 stars with masses $> 10
\;\msun$ among about 380 stars expected in a cluster of stars present
in the interaction region between the expanding HII region and the
molecular cloud in the HII region Sh2-219. \citet{zavagnoetal10} find
around the HII region RCW120 a single massive star (8-10 $\msun$) plus
several low mass stars in the range of 0.8-4 $\msun$. The ages of the
low mass stars is about 50,000 years, compared to an age of 2.5 Myr
for the parent ionizing star determined from the parent star's
photometry and spectral classification \citep{martinsetal10}. This
suggests that the cluster of low mass stars are coming from a second,
presumably triggered generation of stars.  One can also appeal to
other clusters, not necessarily formed by triggered star formation,
that have been studied. \citet{darioetal12} have found that the
initial mass function (IMF) in the Orion Nebula Cluster is similar to
the IMF calculated by \citet{kroupa01} and or \citet{chabrier03} down
to about 0.3 $\msun$, below which it appears to be truncated. Thus,
although \citet{whitworthetal94a} may be correct in that there may be
some preference towards high-mass stars, it does not appear that there
is a significant deviation of the IMF in star-forming regions from the
general interstellar IMF. In general, it seems plausible to assume
that a solar-mass star is not a special case but may be reasonably
expected as a result of triggered star formation.

It is clear that triggered star formation, in bubbles as well as HII
regions, colliding clouds and supernova remnants, is seen.  There
appears to be a correlation between infrared bubbles and star
formation \citep{kendrewetal12}. Many authors
\citep{whitworthetal94a,whitworthetal94b, walch14} have suggested that
massive stars form via triggered star formation and that this can lead
to another generation of triggered stars, leading to so-called
sequential star formation. A further question may be whether the
formation of a solar mass star via triggering is an unusual
circumstance or something that is common. Unfortunately, resolving
this requires knowledge of the importance of triggered and sequential
star formation versus spontaneous star formation.  With a few
assumptions, we can make a best case scenario argument as to the
probability of a given mass star arising from triggered rather than
spontaneous star formation. Let us assume that $N$ stars are born
  at any given time, out of which $N_{WR}$ stars have a mass $> 25\,
  \msun$ and go on to form W-R stars, producing a wind-blown bubble
  with a dense shell. $N_G$ stars have a mass between 0.85 and 1.15
  $\msun$ and constitute solar mass stars. The probability that the
  shell collapses to form stars is $\beta\; (\le 1)$. If it does
  collapse, $N_t$ stars are formed in the swept-up shell due to
  triggering. Then the total number of stars formed as a result of
  triggered star formation in dense shells around W-R stars will be
  $\beta\,N_t\, N_{WR}$. If ${N_G}_t$ solar mass stars are formed due
  to triggered star formation, then the fraction of solar mass stars
  in the shell is ${N_G}_t/N_t$. The total number of solar mass stars
  formed due to collapse of the shell is then given as $(N_{WR}
  \,N_t\, \beta\,{N_G}_t)/N_t$. The fraction of solar mass stars
  formed due to triggering, to the total number of solar mass stars
  formed at any given time, is then $(N_{WR} \,N_t\,
  \beta\,{N_G}_t/(N_t \, N_G))$. This can be considerably simplified
  if we assume that the IMF of the population of triggered stars is
  the same as that of the population of all newly born stars. In that
  case ${N_G}_t/N_t = {N_G}/N$ and the fraction of solar mass stars
  formed by triggering ($F_G$) is:

\be
\label{eq:trig}
 F_G = (N_{WR} \,N_t\, \beta)/N .  
\ee

 We can compute the ratio ${N_{WR}/N}$ from the initial mass function
 (IMF).  Using the Kroupa IMF, where the number of stars between mass
 $M$ and $M + d M$ goes as $\xi M = M^{- \alpha} dM$, with
 $\alpha=2.3$ for $M > 0.5 \msun$ and $\alpha=1.3$ for $0.08 < M < 0.5
 \msun$, the fraction of stars $> 25\,\msun$ is 4.1 $\times
 10^{-3}$. If we assume that gravitational collapse occurs in 10\% of
 the cases ($\beta =0.1$), and that each triggering episode makes a
 100 stars on average, then the fraction of solar mass stars formed by
 triggering is 4.1 $\times\, 10^{-3} \times\,0.1\,\times\,100 \sim
 4\,\%$ of all stars formed at a given time.  The uncertainties
 include what fraction of the shells actually collapse to form stars,
 and the total number of stars produced by triggering, or equivalently
 the total disk mass that collapses to form stars. Since these numbers
 are not well calibrated and could vary by a factor of 2 on either
 end, we estimate conservatively that between 1-16\% of solar mass
 stars could be formed in this manner.

The ratio given in equation \ref{eq:trig} above is equally applicable
to stars of any given mass, since the stellar mass cancels out. This
is due to the assumption that the IMF of triggered star formation is
the same as the IMF for all stars. If this were not the case, then we
would have in equation \ref{eq:trig} an additional factor
${N_G}_t/N_G$, which basically requires knowing the fraction of solar
mass stars due to triggering to the total number of solar mass stars,
or equivalently the IMF in each case. We also note here that only
wind-blown bubbles around W-R stars are being considered. If we were
to take into account bubbles around all main-sequence massive (O and
B) stars, the number of stars formed by triggering could be higher,
although those stars would not be enriched in \altwentysix.

Binaries may modify the conclusion above regarding the number of
Wolf-Rayet stars, and the number of solar mass stars formed by
triggering. It is possible that in a binary, due to more efficient
mass transfer or a higher mass-loss rate, the Wolf-Rayet stage could
be reached at a lower mass. \citet{georgyetal12} have shown that
rotation can also lead to W-R stars forming at a lower mass threshold
of about 20 $\msun$.

We have assumed a single central star to be responsible for the
\altwentysix. However as mentioned earlier, dust around W-R stars has
been seen predominantly in stars that are binaries. Furthermore
massive stars seem to like company - a recent review \citep{sana17}
suggested that 50-70\% may be in binaries, with some surveys
suggesting as high as 90\%. Thus it is quite likely that it was not a
single star but more likely a W-R star with a companion. These
companions are found to usually be other massive O stars. This does
not alter our scenario, and possibly enhances the results, because the
probability of having sufficient \altwentysix to pollute the dense
shell increases, the expectation of dust formation increases, and the
amount of dust formed may be higher.

Can the abundances of other short-lived radionuclides found in the
early solar system also be adequately explained by this process?  Here
we focus on two other species whose early solar system abundances are
commonly thought to be possibly due to late incorporation of fresh
stellar ejecta, namely, $^{36}$Cl and $^{41}$Ca.  One of the problems
in addressing this question is in obtaining the yields of these
species. Although \citet{agm06} claim that the abundance of $^{36}{\rm
  Cl}/^{35}{\rm Cl}$ carried by the W-R wind is sufficient to produce
the value of $^{36}{\rm Cl}/^{35}{\rm Cl} = 1.4 \pm 0.2 \times
10^{-6}$ observed in the initial solar system, this does not seem
obvious from the plots presented in the paper. Both
\citet{gaidosetal09} and \citet{tdd10} find that using the yields
given in the paper, the value is much lower (by orders of magnitude)
than the value reported for CAI's.  The problem is further compounded
by the fact that the initial solar system yield is itself not well
calibrated. Most recently, \citet{tangetal17} have calculated the
initial solar system value from Curious Marie, an aqueously altered
Allende CAI, and found it to be a factor of 10 higher than quoted in
\citet{agm06}, which would further increase the discrepancy. It is
clear that the yields presented in \citet{agm06} would not be able to
satisfy this higher value. Other authors \citep{wasserburgetal11,
  lugaroetal12} have similarly suggested that this isotope is unlikely
to arise from a stellar pollution scenario. This reinforces the
suggestion made previously in the literature, that $^{36}{\rm Cl}$ is
formed primarily as a result of energetic particle irradiation
\citep{gv00, wasserburgetal11}.

The value of $^{41}{\rm Ca}/^{40}{\rm Ca}$ for the initial solar
system was found by \citet{lcsm12} to be 4.2 $ \times 10^{-9}$,
primarily based on two CAIs from the CV chondrite Efremovka. It
appears from the yields given for a 60 $\msun$ star by \citet{agm06}
that this could be easily satisfied in the current model. A crucial
question is when exactly the $^{41}{\rm Ca}$ was emitted, and how long
it took to be injected into the dense shell, given the short half life
of 10$^5$ yr for $^{41}{\rm Ca}$. This requires more information than
is available in current stellar evolution models. \\

\acknowledgements We thank the anonymous referee for a comprehensive
reading of the manuscript, and for their comments and suggestions
which helped to improve this work. This work is supported by a NASA
Emerging Worlds program grant \# NNX15AH70G awarded to the University
of Chicago and a NASA Cosmochemistry grant \#NNX14AI25G awarded to
Clemson University. VVD would like to thank Prof.~Arieh Konigl for a
highly instructive discussion, and for his suggestions. We thank
Prof.~Georges Meynet for providing their group calculations of the
evolution of W-R stars, including the \altwentysix production.

\bibliographystyle{aasjournal} 
\bibliography{references}

\label{lastpage}



\end{document}